\documentclass[prl,floatfix,twocolumn,superscriptaddress,showpacs,preprintnumbers,longbibliography,nofootinbib]{revtex4-1}

\usepackage[utf8]{inputenc} %
\usepackage{amssymb,amsfonts,amsmath}
\usepackage{float} 
\usepackage{amsthm}
\usepackage{enumitem}
\usepackage{bbold}

\usepackage{hyperref} 
\hypersetup{colorlinks=true, urlcolor=blue, citecolor=blue,linkcolor=blue}
\usepackage[all]{hypcap} 

\usepackage{graphicx}
\usepackage{dcolumn}
\usepackage{bm}
\usepackage{psfrag}
\usepackage{epsfig}
\usepackage{color}
\usepackage{bbm}
\usepackage[FIGTOPCAP,raggedright,nooneline]{subfigure}
\usepackage{verbatim}
\usepackage{upgreek} 

\usepackage{tikz} 

\newcommand{\T}[1]{\text{#1}}
\newcommand{\bra}[1]{\langle #1 |}
\newcommand{\ket}[1]{| #1 \rangle}

\newcommand{\ignore}[1]{}

\newcommand{\eq}{Eq.\,}

\newcommand{\fig}{Fig.\,}
\newcommand{\figs}{Figs.\,}
\newcommand{\cf} {cf.~}

\newcommand{\rref} {Ref.\,}

\begin{document}

		\title{	
		Stationary quantum correlations in a system with mean-field $\mathcal{PT}$ symmetry }
		
		\author{Federico Roccati }
		\affiliation{Universit$\grave{a}$  degli Studi di Palermo, Dipartimento di Fisica e Chimica -- Emilio Segr$\grave{e}$, via Archirafi 36, I-90123 Palermo, Italy}
	
		\author{Salvatore Lorenzo }
		\affiliation{Universit$\grave{a}$  degli Studi di Palermo, Dipartimento di Fisica e Chimica -- Emilio Segr$\grave{e}$, via Archirafi 36, I-90123 Palermo, Italy}
	
		\author{G.~Massimo Palma}
		\affiliation{Universit$\grave{a}$  degli Studi di Palermo, Dipartimento di Fisica e Chimica -- Emilio Segr$\grave{e}$, via Archirafi 36, I-90123 Palermo, Italy}
		\affiliation{NEST, Istituto Nanoscienze-CNR, Piazza S. Silvestro 12, 56127 Pisa, Italy}
	
		\author{Francesco Ciccarello}
		\affiliation{Universit$\grave{a}$  degli Studi di Palermo, Dipartimento di Fisica e Chimica -- Emilio Segr$\grave{e}$, via Archirafi 36, I-90123 Palermo, Italy}
		\affiliation{NEST, Istituto Nanoscienze-CNR, Piazza S. Silvestro 12, 56127 Pisa, Italy}
	
		\date{\today}
	
		\begin{abstract}
			A pair of coupled quantum harmonic oscillators, one subject to a gain one to a loss, is a paradigmatic setup to implement PT-symmetric, non-Hermitian Hamiltonians in that one such Hamiltonian governs the {\it mean}-field dynamics for equal gain and loss strengths. Through a full quantum description (so as to account for quantum noise) here is shown that when the system starts in any two-mode coherent state, including vacuum, there appear quantum correlations (QCs) without entanglement, as measured by the Gaussian discord. When the loss rate is above a threshold, once generated QCs no more decay. This occurs in a wide region of parameters, significantly larger than that where the full quantum dynamics is stable. For equal gain and loss rates, in particular, QCs decay in the exact phase (including the exceptional point) and are stable in the broken phase.
		\end{abstract}
	
		\maketitle
	
			\section{Introduction}
			The 1998 discovery of non-Hermitian Hamiltonians that yet have real eigenvalues \cite{benderPRL1998} introduced a new class of dynamics. This has fueled widespread attention at a fundamental level as well as in connection with a number of potentially appealing applications \cite{el-ganainyNP2018,fengNP2017,longhiEPL2017}. Much of this interest is motivated by the possibility that dynamics effectively governed by such non-Hermitian Hamiltonians are experimentally observable, especially in optics
			\cite{ruterNP2010,regensburgerN2012,pengNP2014}. A prototypical one is a gain-loss system (see \fig1) comprising a pair of mutually-coupled modes (oscillators), one subject to a gain with rate $\gamma_{G}$ and one to a loss with rate $\gamma_{L}$ such that $\gamma_{G}=\gamma_{L}=\gamma$. The modes' mean field, represented by a two-dimensional vector, evolves according to a Schrodinger-like equation with a 2$\times$2 non-Hermitian Hamiltonian $\cal H$ that enjoys parity-time (PT) symmetry. As such, this has two real eigenvalues in the so called exact phase ($\gamma$ smaller than the inter-mode coupling strength $g$), while in the broken phase $\gamma>g$ there are two complex eigenvalues that coalesce at the exceptional point $\gamma=g$.

			To derive these effective non-Hermitian Hamiltonians, it is enough to invoke Maxwell's equations. In this sense, such dynamics are essentially classical. A somewhat more fundamental way to see this is modeling the two waveguides as {\it quantum} harmonic oscillators whose joint dynamics obeys a Lindblad master equation. 
			This features a unitary term, corresponding to a beam-splitter-like interaction Hamiltonian between the oscillators \cite{haroche2006}, plus two local ``dissipators'' each with an associated jump operator: one for the gain (incoherent pump), one for the loss. It turns out that the corresponding evolution of the pair of {\it mean} fields is governed by a PT-symmetric, non-Hermitian Hamiltonian. In light of the correspondence principle, such Hamiltonians thus rule dynamics where quantum noise is negligible or neglected. 
			
			While current years are witnessing a burst of interest for quantum technologies, to date only a relatively small number of works investigated genuinely quantum properties of  $\mathcal{PT}$ symmetric systems or related \cite{schomerusPRL2010,yooPRA2011,agarwalPRA2012,longhiOL2018,vashahri-ghamsariPRA2017,vashahri-ghamsariPRA2019}. 
			In particular, their potential to exhibit novel phenomena, and ensuing possible applications, that rely on the very quantum nature of the field is yet largely unexplored. A major issue in this respect is the intrinsic noise unavoidably introduced by the local gain and loss \cite{kepesidisNJP2016,lauNC2018,zhangAQ2018}, 
			which does not bode well for occurrence of quantum coherent phenomena, especially entanglement \cite{nielsen2010}, spotlighting a substantial difference from typical quantum optics settings used for quantum information processing applications \cite{braunsteinRMP2005} (this motivated an alternative dissipationless implementation of non-Hermitian Hamiltonians \cite{wangPRA2019}). Intense research activity over the last decade, however, has shown in various ways the existence of ``cheap'' quantum resources that put mild constraints on the necessary amount of quantum coherence. Among these is a form of exploitable quantum correlations (QCs) that can occur even in absence of entanglement. First discovered in 2001 \cite{ollivierPRL2001,hendersonJPAMG2001}, this extended paradigm of QCs
			\begin{figure}
				\centering
				\includegraphics[width=5cm]{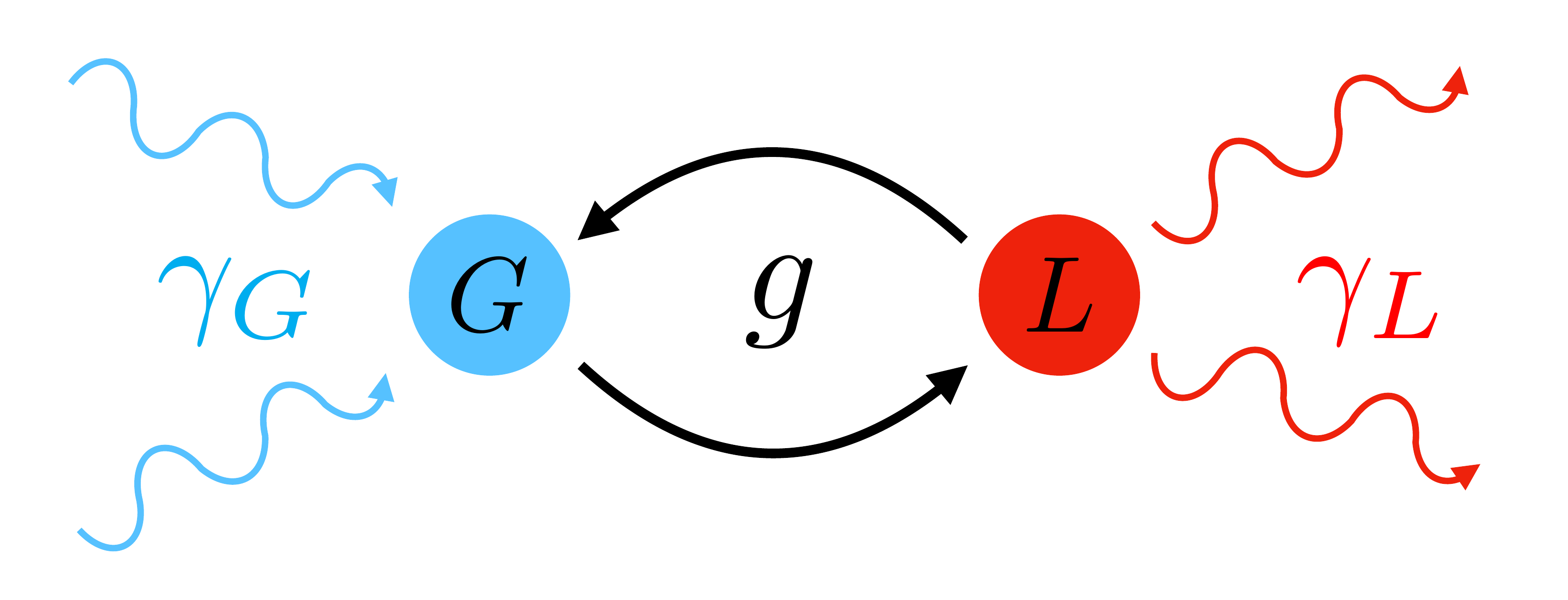}
				\caption{Sketch of the gain-loss system. A pair of quantum harmonic oscillators $G$ and $L$ (modes) are directly coupled with coupling strength $g$. Additionally, mode $G$ is subject to a local gain while a loss acts on $L$ (with characteristic rates $\gamma_{G}$ and $\gamma_{L }$, respectively). The modes start in a coherent state $\ket{\alpha_{G}}\!\otimes\!\ket{\alpha_{L}}$, sharing no initial correlations (the $\alpha$'s are fully arbitrary). While entanglement never shows up, quantum correlations are nevertheless created and, in a wide region of parameters, no more decay.}
			\end{figure}
			has received huge attention \cite{modiRMP2012}, both theoretical and experimental, after it was first realized that it can be harnessed for carrying out quantum algorithms \cite{dattaPRL2008}. Remarkably, a recent work reported first experimental detection of such a form of QCs \cite{caoAQ2019} in {\it anti}-PT symmetric system featuring similarities with the setup in \fig1.

			This preliminary report investigates appearance of QCs in the case study of the gain-loss setup in \fig1 when each oscillator starts in a coherent state (even vacuum) and no correlations of any sort are initially present. 
			Besides their routine production in the lab, the choice of coherent states allows to connect with standard classical optics setups in that one such state has minimum quantum uncertainty \cite{loudon2000}. It will be shown that, while no entanglement is produced, the system develops QCs measured by the Gaussian discord. Moreover, when the loss rate is above a threshold, the generated QCs no more decay.

			\section{Gain-loss system} 
			We consider two quantum harmonic oscillators $G$ and $L$ (see \fig1), whose joint state (described by the density matrix $\rho$) evolves in time according to the Lindblad master equation (we use units such that $\hbar=1$)
			\begin{equation}
			\dot\rho  =  - i[H,\rho] +2\gamma_{L}\, \mathcal D[\hat a_L]\rho +2\gamma_{G} \,\mathcal D[\hat a_G^\dag]\rho \label{ME}
			\end{equation}
			with $\mathcal D[\hat A]\rho=\hat A \rho A^\dagger - \tfrac{1}{2}(\hat A^\dag \hat A \rho+\rho\hat A^\dag\hat A)$ and
			\begin{equation}
			H = g\, ( a_{L }^\dagger a_{G}+a_{L }a_{G}^\dagger ) \,.\label{H}
			\end{equation}
			Here, $\hat a_{G(L)}$ and $\hat a^\dag_{G(L)}$ are usual bosonic ladder operators such that $[\hat a_{G(L)},\hat a^\dag_{G(L)}]=1$. We implicitly assumed a bare Hamiltonian $\hat H_0=\sum_{n=G,L}\omega \,\hat a^\dag_{n}\hat a_{n}$ for the two modes such that ME \eqref{ME} holds in the interaction picture with respect to $\hat H_0$. 
			The interaction Hamiltonian \eqref{H} describes a coherent energy exchange at rate $g$ between the modes. In addition, each mode interacts incoherently with a local environment: the one on $G$ pumps energy with characteristic rate $\gamma_G$ (gain) while that on $L$ absorbs energy with rate $\gamma_L$ (loss). Formally, the irreversible action of either environment on the system is described by a local jump operator [see \eq\eqref{ME}]: this is a creation operator for the gain and a destruction operator for the loss ($\hat a_G^\dag$ and $\hat a_L$, respectively). Analogous jump operators to describe gain and loss of a PT-symmetric system were used for instance in \rref\cite{dastPRA2014}.

			\section{Mean-field dynamics}
			The expectation value of an operator $\hat O$ evolves in time according to $\tfrac{\T{d}}{\T{d}t}{\langle \hat O\rangle}=\langle \hat O \dot \rho\rangle$. Replacing $\dot \rho$ with \eqref{ME} yields that the time evolution of the two-dimensional complex vector $\psi=(\langle \hat{a}_L\rangle, \langle \hat{a}_G\rangle)^T$, which describes the mean field, is governed by the Schrodinger-like equation $i \dot{\psi} =  \mathcal H \psi$ with 
			\begin{align}\label{evolMeanVal}
			\mathcal H=\left(
			\begin{array}{cc}
			-i\gamma_{L } & g \\
			g & i\gamma_{G} \\
			\end{array}
			\right) \,.
			\end{align}
			The non-Hermitian matrix $\cal H$ generally has two complex eigenvalues with associated non-orthogonal eigenstates. For $\gamma_L=\gamma_G=\gamma$, $\cal H$ has PT-symmetry, i.e., it is invariant under the swap $G\leftrightarrow L$ plus time reversal. In this case, the $\cal H$ eigenvalues are given by $\varepsilon=\pm \sqrt{g^2-\gamma^2}$: these are real in the exact PT phase $\gamma<g$ and complex in the broken phase $\gamma>g$, coalescing at the exceptional point (EP) $\gamma=g$ where the corresponding eigenstates become parallel.

			\section{Second-moment dynamics}
			The field has associated quantum uncertainty described by a covariance $4\times 4$ matrix, whose entries are the expectation values of all possible products of two ladder operators.
			It is however convenient to describe each mode in terms of its quadratures $(\hat x_n, \hat p_n)$ with $\hat x_n=\tfrac{1}{\sqrt{2}}(\hat a_n+\hat a_n^\dag)$ and $\hat p_n=\tfrac{i}{\sqrt{2}} (\hat a_n^\dagger - \hat a_n)$, and define the covariance matrix $\sigma_{ij} = \langle \hat X_i \hat   X_j+\hat X_j \hat X_i\rangle$ with $\hat X_i=( \hat x_L, \hat p_L,\hat x_G, \hat p_G)$ \cite{gardiner2004}. The block-structure of $\sigma$, which has real entries, reads
			\begin{equation}\label{blocks}
			\sigma=\left(
			\begin{array}{cc}
			L & C \\
			C^T & G \\
			\end{array}
			\right)\,,
			\end{equation}
			thus the diagonal $2\times 2$ blocks describe uncertainties affecting the local fields, while the off-diagonal block $C$ accounts for $G$-$L$ cross-correlations. Following a standard recipe \cite{gardiner2004}, ME \eqref{ME} entails the equation of motion for the covariance matrix 
			\begin{equation}\label{eqforsigma}
			\dot{\sigma} = Y\sigma +\sigma\, Y^{T} + 4 D 
			\end{equation}
			with 
			\begin{equation}
			Y =\left(
			\begin{array}{cccc}
			-\gamma_{L }  & 0 & 0 & g \\
			0 & -\gamma_{L }  & -g & 0 \\
			0 & g & \gamma_{G}  & 0 \\
			-g & 0 & 0 & \gamma_{G} \\
			\end{array}
			\right)\label{Y}
			\end{equation} 
			and
			$D=\tfrac{1}{2}\,{\rm diag}(\gamma_{L},\gamma_{L},\gamma_{G},\gamma_{G})\label{D}$. 
			The dynamics studied in this work involve solely two-mode Gaussian states. One such state is fully specified by the mean-field vector $\psi$ and covariance matrix $\sigma$.
			
			\section{Classical and quantum correlations}
			\begin{figure}
				\centering
				\includegraphics[width=8.7cm]{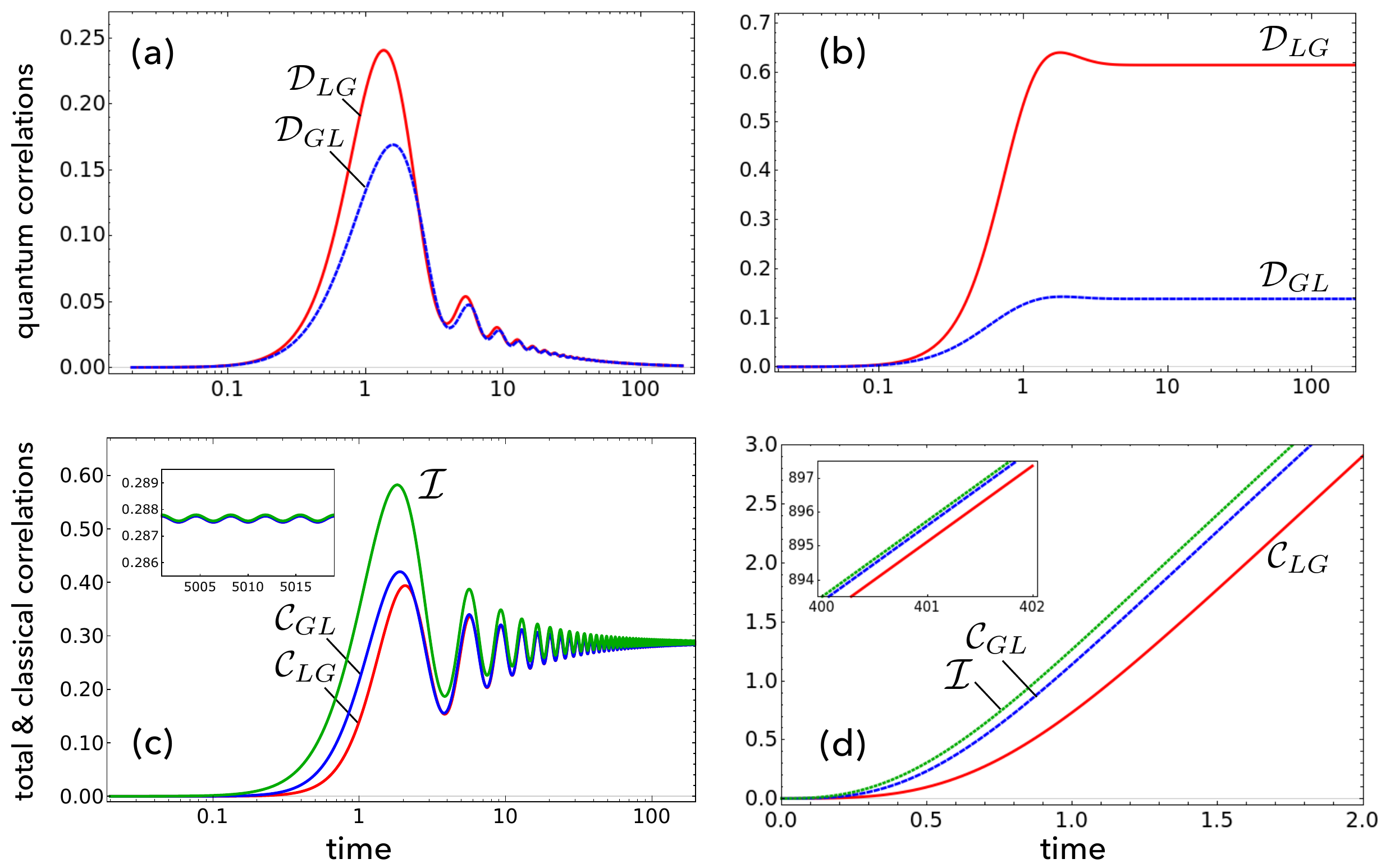}
				\caption{Total, classical and quantum correlations against time (in units of $g^{-1}$) for $\gamma_G=\gamma_{L}=g/2$ [panels (a) and (c)], and $\gamma_G=\gamma_{L}=3g/2$ [(b) and (d)]. (a) and (b): $\mathcal D_{GL}$ (blue dashed line) and $\mathcal D_{LG}$ (red solid). (c) and (d): $\mathcal I$ (green dotted), $\mathcal C_{GL}$ (blue dashed) and $\mathcal C_{LG}$ (red solid). Insets in (c) and (d) show the long-time behavior. All plots are in log-lin scale.}
			\end{figure}
			A popular quantifier of correlations is the mutual information $\cal I$ \cite{cover2006}.  For a bipartite system, this is the difference between the sum of local entropies and the entropy of the joint system, $\mathcal I= S(G)+S(L)-S({GL})$. The quantum version reads \cite{nielsen2010} $\mathcal I= S(\rho_{G})+S(\rho_{L})-S(\rho)$, with $\rho_{G}={\rm Tr}_{L}\rho$ the marginal describing the state of subsystem $G$ (analogously for $\rho_L$) while $S(\varrho)=-\T{Tr}(\varrho\log\varrho)$ is the standard Von Neumann entropy of a quantum state $\varrho$. The mutual information $\cal I$ captures the entire amount of correlations (both classical and quantum). Indeed, $\mathcal I=0$ if and only if $\rho=\rho_G\otimes \rho_L$ (product state). For classical systems, $\cal I$ can be equivalently expressed as $\mathcal I=\mathcal C_{GL}=S(L)-S(L|G)$, with $S(L|G)$ the conditional entropy associated with a local measurement on $G$ (changing the roles of $G$ and $L$ does not affect the result, that is $\mathcal{I}=\mathcal C_{GL}=\mathcal C_{LG}$).  For quantum systems, this equivalence generally does {\it not} hold \cite{ollivierPRL2001,hendersonJPAMG2001}: the mismatch between the two definitions is measured by quantum {\it discord}. This is formally defined as follows. One first formulates the quantum version of $\mathcal C_{GL}$ as 
			\begin{equation}
			\mathcal C_{GL}=S(\rho_{L})-\underset{\hat G_k}{\rm min}\,\,\sum_k p_k S(\rho_{L|k})\,.\label{cc}
			\end{equation}
			Here, $\{\hat G_k\}$ describes a local projective measurement on subsystem $G$ with possible outcomes indexed by $k$ and occuring with probability $p_k={\rm Tr}( \rho\, \hat G_k)$. Thus, a measurement on $G$ yielding the result $k$ collapses the joint state $\rho$ onto $\rho_{L|k}={\rm Tr}_G( \rho\, \hat G_k)/p_k$. Given a quantum system, there exist infinite sets $\{\hat G_k\}$ (for a spin-1/2 particle, e.g., there are as many as the axes along which spin can be measured): the infimum in \eq\eqref{cc} is over all possible sets $\{G_k\}$. Quantum discord is defined as \cite{modiRMP2012}
			\begin{equation}\label{quantumdiscord}
			\mathcal D_{GL}=\mathcal I-\mathcal C_{GL}\,,
			\end{equation}
			fulfilling $\mathcal D_{GL}\ge 0$. Notably, unlike the classical case, in general $\mathcal C_{GL}\neq \mathcal C_{LG}$ for quantum systems. As a consequence, discord is generally {\it asymmetric}: $\mathcal D_{GL}$ can differ from $\mathcal D_{LG}$. Recalling that $\cal I$ measures the entire amount of correlations, \eqref{cc} is interpreted as the amount of classical correlations and discord \eqref{quantumdiscord} as the measure of QCs. Notably, any entangled state has non-zero discord. Yet, the converse does not hold: there exist non-entangled states that are ``discordant'' \cite{modiRMP2012}. Thereby, discord can detect QCs that do {\it not} give rise to entanglement: in this sense, it can be seen as the most general measure of QCs.
			
			The explicit calculation of \eqref{cc} and \eqref{quantumdiscord} is typically tough, even more so for
			infinite-dimensional systems such as the two modes in \fig1. Yet, in the relevant case of Gaussian states, one can restrict \eqref{quantumdiscord} to {\it Gaussian} local measurements \cite{pirandolaPRL2014}. The resulting measure of QCs takes the name of {\it Gaussian discord}. Its analytical form, as well as that of $\mathcal I$ and classical correlations \eqref{cc}, is known for {\it any} Gaussian state as an explicit function of the covariance matrix \eqref{blocks} \cite{giordaPRL2010,adessoPRL2010}.
			
			As mentioned, discord detects QCs more general than entanglement. For Gaussian discord, this is condensed in a simple property: Gaussian states such that $\mathcal D>1$ are entangled, while for $0<\mathcal D<1$ entanglement is zero \cite{adessoPRL2010}.

			\section{Dynamics of quantum correlations}
			We study creation QCs for the gain-loss system in \fig1 when each oscillator $n=G,L$ starts in a coherent state $\ket{\alpha_{n}}=e^{(\alpha\hat a_n^\dag-\alpha^*\hat a_n)}\ket{0}$ of unspecified amplitude $\alpha_n$ (here $\ket{0}$ is the vacuum state). 
			The joint initial density matrix thus reads $\rho_0=\ket{\alpha_{G}}\!\bra{\alpha_{G}}\otimes\ket{\alpha_{L}}\!\bra{\alpha_{L}}$, which is a product state featuring zero $G$-$L$ correlations (even classical ones). The corresponding covariance matrix is simply the 4$\times$4 identity, being thus independent of the $\alpha_n$'s. To compute the dynamics of correlations, we evolve the covariance matrix through \eq\eqref{eqforsigma} from which we infer the explicit time-dependence of $\mathcal I$, $\mathcal C$ and $\mathcal D$. 
			
			In \fig2(a), we set $\gamma_G=\gamma_{L}=g/2$ (whose corresponding point in the parameters space lies on the exact-phase segment) and plot the time behavior of $\mathcal{D}_{GL}$ as well as $\mathcal{D}_{LG}$. They both grow from zero, reach a maximum and fade away altogether after a slow decay. Note that (generally) $\mathcal{D}_{GL}(t)\neq\mathcal{D}_{LG}(t)$, which can be expected from the intrinsic asymmetry of the gain-loss system. Moreover, discord remains at any time below the entanglement threshold $\mathcal D=1$. Therefore, despite entanglement is identically zero, QCs are generated in the transient although they disappear at long times. A similar behavior occurs for mutual information and classical correlations, except that these all saturate to the same finite value [see \fig2(c)]. The three curves $\mathcal I(t)$, $\mathcal C_{GL}(t)$ and $\mathcal C_{LG}(t)$ are disjoint at intermediate times, confirming creation of quantum correlations in the transient [\cf\eq\eqref{quantumdiscord}].
			A different choice of parameters is made in \figs2(b) and (d), where we set $\gamma_G=\gamma_{L}=3g/2$ (lying on the broken phase). Again [see \fig2(b)] both $\mathcal{D}_{GL}$ and $\mathcal{D}_{LG}$ rise up without ever trespassing $\mathcal D=1$. Yet, at variance with \fig2(a), now discord eventually saturates to a {\it finite} value. Thus, in this case, not only is discord created but this survives at long times, when it becomes stationary. In stark contrast, mutual information and classical correlations [see \fig2(d)] all undergo a continuous growth, an unstable behavior that can be ascribed to the presence of an active element (the gain). At long times, $\mathcal I(t)$, $\mathcal C_{GL}(t)$ and $\mathcal C_{LG}(t)$ run parallel, witnessing establishment of stable QCs. 
			\begin{figure}
				\centering
				\includegraphics[width=8.5cm]{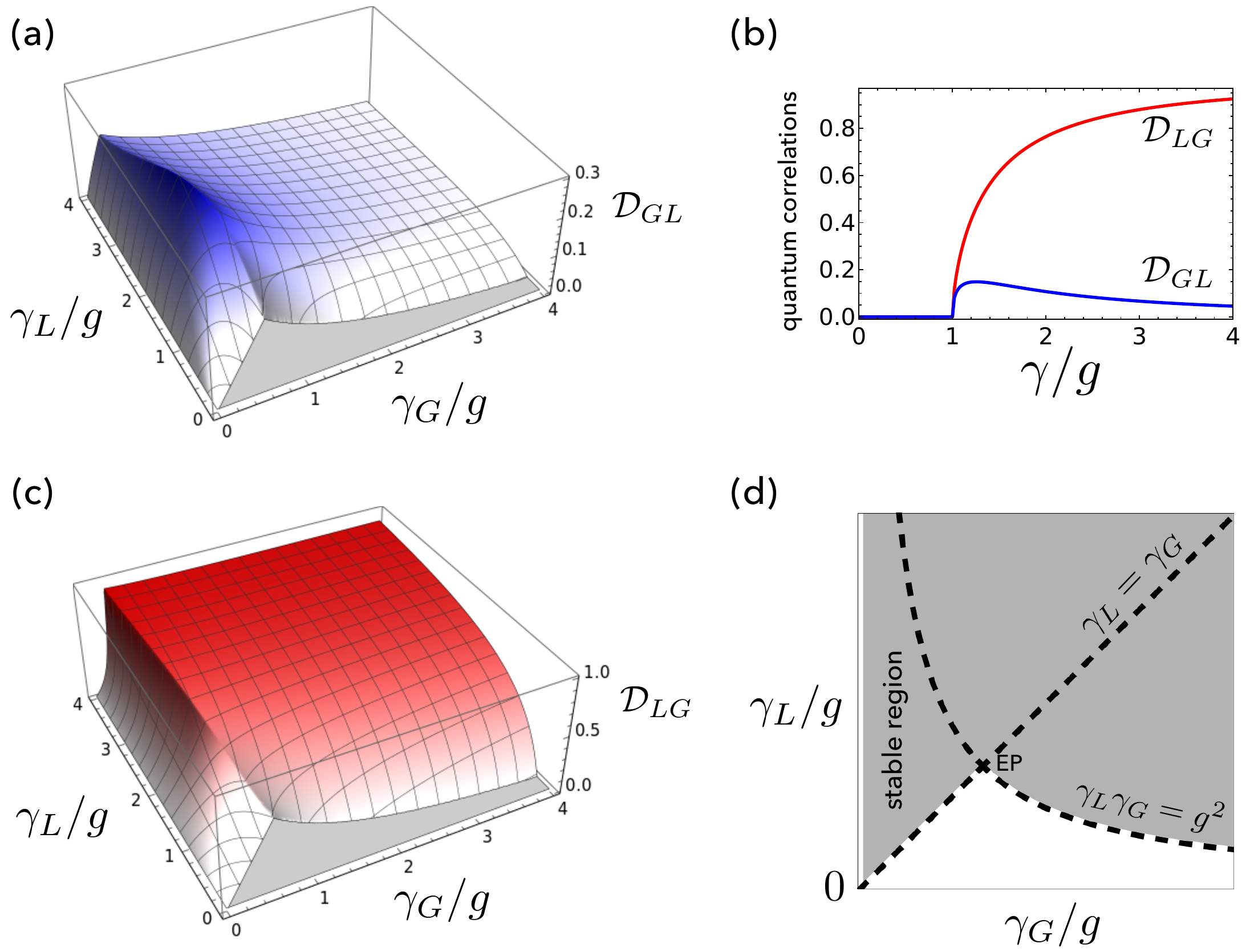}
				\caption{(a) and (c): Long-time QCs, as measured by $\mathcal{D}_{GL}$ (a) and $\mathcal{D}_{LG}$ (c), on the plane $\gamma_{G}$ -- $\gamma_{L }$ (in units of $g$). (b): Behavior of $\mathcal{D}_{LG}$ (red line) and $\mathcal{D}_{GL}$ (blue) along the line $\gamma_{L}=\gamma_{G}=\gamma$, where mean-field Hamiltonian \eqref{evolMeanVal} is PT-symmetric. The segment $\gamma/g<1$ is the exact phase and the semi-infinite line $\gamma/g>1$ the broken phase, with the exceptional point $\gamma=g$ at the boundary of the two phases. (d): Schematic diagram showing the region of finite asymptotic QCs (grey) and zero discord (white); note that the latter includes the vertical line $\gamma_G=0$, the exact-phase segment and the exceptional point ``EP''. The dynamics is fully stable ($\mathcal H$ and $\sigma$ converging to finite values) in the region bounded by the $\gamma_G=0$ line, the PT-symmetry line $\gamma_{L}=\gamma_{G}$ and the hyperbola $\gamma_{G}\gamma_{L }=g^2$.}
			\end{figure}
			
			Creation of quantum correlations that eventually decay [as in \fig2(a)] or survive [as in \fig2(b)] are not specific to the instances considered in \fig2. The asymptotic value  of $\mathcal{D}_{GL}$ and $\mathcal{D}_{LG}$ is shown in \figs3(a) and (c), respectively, as a function of $\gamma_G$ and $\gamma_L$ (in units of $g$). As is apparent in both figures, in addition to $\gamma_G=0$ no stationary discord occurs in the region bounded by the $\gamma_L=0$ line, the PT-symmetry line $\gamma_{L }=\gamma_{G}=\gamma$ and the hyperbola $\gamma_{G}\gamma_{L }=g^2$. This includes the exceptional point (EP) $\gamma_{G}=\gamma_{L }=g$, at which the PT-symmetry line and hyperbola intersect, as well as the PT-exact-phase segment. Note that, for fixed (non-zero) gain, stationary QCs are finite when loss $\gamma_{L }$ is above the threshold $\gamma_{\rm th}=\gamma_G$ for $\gamma_{G}\le g$, and $\gamma_{\rm th}=g^2/\gamma_{G}$ for $\gamma_{G}> g$.  Along the PT-simmetry line in particular, as better detailed in \fig3(b), asymptotic QCs identically vanish in the exact phase including the EP and are non-zero above the EP ($\gamma>g$). In the last range, $\mathcal{D}_{GL}$ and $\mathcal{D}_{LG}$ behave differently in that the former reaches a maximum and then slowly decreases with $\gamma$, while the latter exhibits a monotonic growth until it saturates to the maximum value compatible with zero entanglement, $\mathcal{D}_{LG}=1$ (not shown). In both \figs2(a) and 2(c), note that non-zero QCs occur in the region $\gamma_{L}>\gamma_G$ below the hyperbola $\gamma_{G}\gamma_{L }=g^2$. This region, as easily checked by inspection of Lyapunov exponents, is the one where the dynamics is fully stable (both in the mean field and covariance matrix). Clearly, QCs are stable well beyond this region, confirming the outcomes for $\gamma_L=3\gamma_G/2$ (broken phase) reported in \fig2(b).
			The behavior on the entire parameters space is summarized in the schematic diagram in \fig3(d).
			
			\section{Physical interpretation}

			The mechanism responsible for QCs creation without entanglement can be understood through the following qualitative argument.
			The coupling Hamiltonian \eqref{H} acts on the modes just like a beam splitter. A beam splitter transforms a two-mode coherent state as $\ket{\alpha_G}\!\bra{\alpha_G}\otimes \ket{\alpha_L}\!\bra{\alpha_L}\rightarrow \ket{\tilde\alpha_G}\!\bra{\tilde\alpha_G}\otimes \ket{\tilde\alpha_L}\!\bra{\tilde\alpha_L}$ \cite{haroche2006}. Hence, in particular, it cannot create entanglement	\cite{kimPRA2002}. Being local non-unitary channels, no aid to establish entanglement is expected from the additional presence of loss and gain. Indeed, entanglement is never generated in our dynamics. When it comes to quantum discord, however, local non-unitary channels can be beneficial \cite{modiRMP2012}. For instance, it is well assessed \cite{ciccarelloPRA2012a,huPRA2012,streltsovPRL2011,ciccarelloPRA2012} that a local gain or loss can create QCs starting from a state featuring only classical correlations (an impossible process for entanglement), which was experimentally confirmed \cite{madsenPRL2012}.
			A loss transforms a coherent state into one of smaller amplitude \cite{haroche2006}, $\ket{\alpha}\rightarrow  \ket{\eta\alpha}$ with $\eta<1$, until this reduces to the vacuum state at long enough times. 
			This entails that the state's purity is unaffected, ${\rm Tr}(\ket{\eta \alpha}\!\bra{\eta \alpha})^2=1$ for any $\eta$. In contrast, the gain turns a coherent state into a {\it mixture} \cite{scheelEPL2018}, $\ket{\alpha}\!\bra{\alpha}\rightarrow \int d^2 \alpha'P(\alpha') \ket{\alpha'}\!\bra{\alpha'}$ with $P(\alpha')\ge 0$ (purity diminished). Now, consider a product of coherent states $\ket{\alpha_G}\!\bra{\alpha_G}\otimes \ket{\alpha_L}\!\bra{\alpha_L}$. A gain followed by a beam splitter transform this into $\int d^2\tilde\alpha'_G P(\tilde\alpha'_G) \ket{\tilde\alpha'_G}\!\bra{\tilde \alpha'_G}\otimes\ket{\tilde\alpha'_L}\!\bra{\tilde \alpha'_L}$. Although not entangled, one such state is generally discordant, which is a consequence of the well-known fact that coherent states form a {\it non}-orthogonal basis \cite{korolkovaRPP2019}. Note that the successive application of a loss will have the effect of making $\ket{\tilde\alpha'_L}\!\bra{\tilde \alpha'_L}$ closer to the vacuum, thus shrinking correlations. Hence, gain tends to create discord and loss to destroy it. A balance between the two can occur, which qualitatively shows the possibility that once created QCs remain stable. It can be checked that, if the gain in \fig1 is replaced by a loss, no discord is produced at any time. On the other hand, replacing the loss in \fig1 with a gain yields QCs in the transient that yet {\it always} decay eventually. The simultaneous presence of gain and loss thus appears essential for stabilizing discord.

			\section{Conclusions}
			We considered a paradigmatic setup to observe PT-symmetric physics, a pair of coupled oscillators subject to local gain and loss, whose dynamics was described through a full quantum treatment (beyond mean field). This leads to the prediction that, starting from an initial pair of uncorrelated coherent states (including vacuum), although entanglement never shows up at any time the system develops QCs measured by the Gaussian discord. When the loss rate is above a threshold, once established QCs no more decay, which in particular occurs in the PT-broken phase. QCs, instead, do not survive in the PT-exact phase. 
			
			As noted above, a gain channel introduces {\it mixedness} (reduced coherence). This particular feature usually appears as a major limitation to harnessing PT-symmetric systems for quantum optics and quantum information applications \cite{scheelEPL2018}, In contrast, in the present dynamics the gain mixedness is just the key resource for creating QCs in the form of discord (along with the coupling). Mixedness is indeed essential to get discord in absence of entanglement \cite{modiRMP2012}. Some readers might object that a mixture of two-mode coherent states, such as those featuring discord here, is a fully classical state according to the longstanding notion of quantumness in quantum optics. In this respect, it has been thoroughly clarified 	\cite{ferraroPRL2012} that the last way to distinguish quantum from classical states {\it differs} from the notion of quantumness underpinning discord, the two criteria almost never coinciding (see also \rref\cite{korolkovaRPP2019}). Plenty of evidence was gathered, even experimentally, that QCs of non-entangled states can be exploited for a number of applications 
			\cite{adessoJPAMT2016,streltsov2015}. Among others: information encoding \cite{guNP2012}, remote-state preparation \cite{dakicNP2012}, entanglement activation \cite{pianiPRL2011, adessoPRL2014, streltsovPRL2011a, croalPRL2015}, entanglement distribution \cite{chuanPRL2012,peuntingerPRL2013, vollmerPRL2013, fedrizziPRL2013}, quantum metrology and sensing \cite{girolamiPRL2014}, 
			The findings presented here thus suggest that, besides mean-field dynamics, dissipative PT-symmetric systems could find applications in quantum technologies.

			\section{Acknowledgements}
			We acknowledge fruitful discussions with M.~Genoni, M.~Paternostro and T.~Tufarelli.
			
			\bibliography{RLPC_v5_arXiv}
			\bibliographystyle{apsrev4-1}

\end{document}